\documentclass[twocolumn,showpacs,preprintnumbers,amsmath,amssymb]{revtex4}

\usepackage{graphicx}% Include figure files
\usepackage{dcolumn}% Align table columns on decimal point
\usepackage{bm}% bold math

\begin{document}

\preprint{APS/123-QED}

\title{N-body decomposition of bipartite networks}


\author{R. Lambiotte}
\email{Renaud.Lambiotte@ulg.ac.be}

\author{M. Ausloos}
\email{Marcel.Ausloos@ulg.ac.be}

\affiliation{%
SUPRATECS, Universit\'e de Li\`ege, B5 Sart-Tilman, B-4000 Li\`ege, Belgium
}%

\date{09/07/2005}

\begin{abstract}

In this paper, we present a method to project co-authorship networks, that accounts in detail for the geometrical structure of scientists collaborations. By restricting the scope to 3-body interactions, we focus on the number of triangles in the system, and show the importance of multi-scientists (more than 2) collaborations in the social network. This motivates the introduction of generalized networks, where basic connections are not binary, but involve arbitrary number of components. We focus on the 3-body case, and study numerically the percolation transition.
   
\end{abstract}


\maketitle

% main text
\section{Introduction}

\begin{figure}
\hspace{1.8cm}
\includegraphics[width=3.00in]{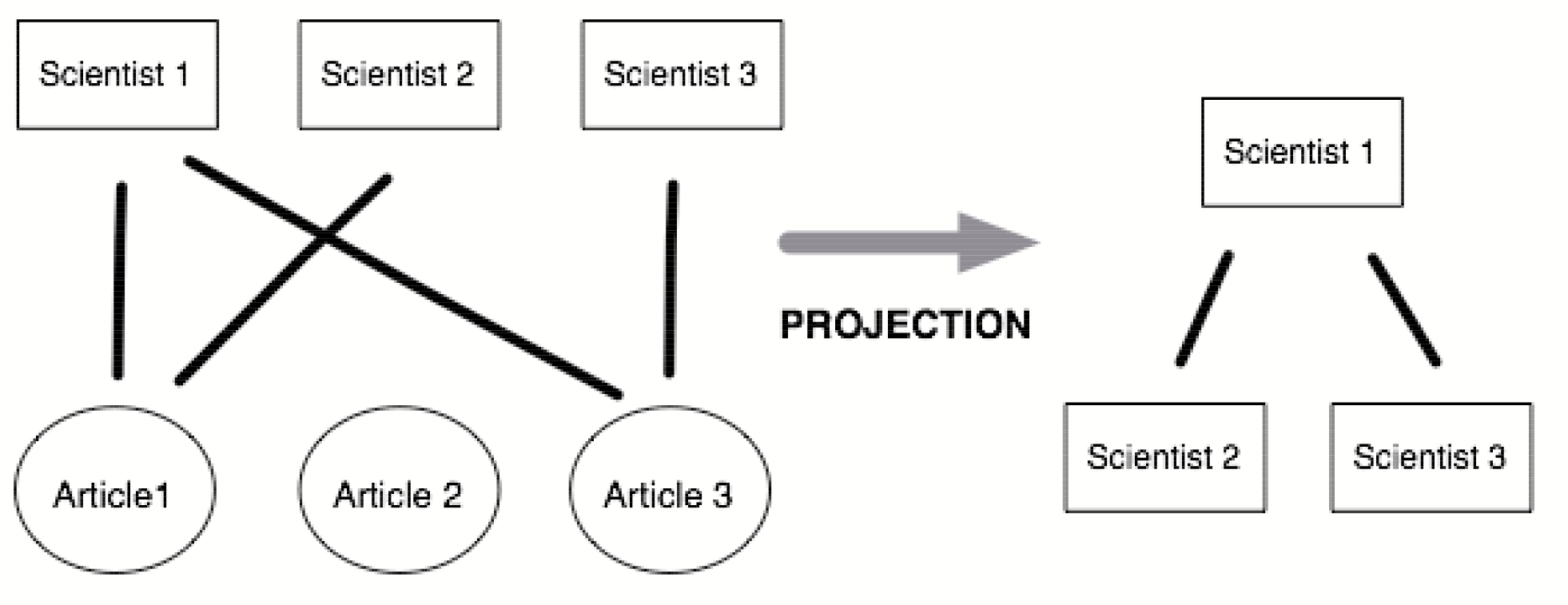}% Here is how to import EPS art
\caption{\label{projection} Usual projection method of the bipartite graph on a unipartite scientists graph.}
\end{figure}

It is well-known in statistical physics that  N-body correlations have to be carefully described in order to characterize statistical properties of complex systems. 
For instance, in the case of the Liouville equation for Hamiltonian dynamics, this problem is at the heart of the derivation of the reduced BBGKY hierarchy, thereby leading to the Boltzmann and Enskog theories for fluids \cite{balescu}. 
In this line of though, it is primordial to discriminate N-body correlations that are due to intrinsic N-body interactions, from those that merely develop from  lower order interactions. This issue is directly related to a well-known problem in complex network theory, i.e. the "projection" of bipartite networks onto simplified structures. As a paradigm for such systems, people usually consider co-authorship networks \cite{newman}, namely networks composed by two kinds of nodes,  e.g. the scientists and the articles, with links running 
between scientists and the papers they wrote.
In that case, the usual
projection method \cite{newman2} consists in focusing e.g. on the scientist nodes, and in drawing a link between them if they co-authored a common paper (see Fig.\ref{projection}). As a result, the projected system is a unipartite network of scientists,  that characterizes the community structure of science collaborations. Such studies have been very active recently, due to their complex social structure \cite{newman3},  to the ubiquity of such bipartite networks in complex systems \cite{bara} \cite{ramasco}, and to the large databases available. 

A standard quantity of interest in order to characterize the structure of the projected network is the clustering coefficient \cite{watts}, which measures network "transitivity", namely the probability that two  scientist's co-authors have themselves coauthored a paper. In topological terms, it is a measure of the density of triangles in a network, a triangle being formed every time two of one's collaborators collaborate with each other. This coefficient is usually very high in systems where sociological cliques develop \cite{eurovision}. However, 
part of the clustering in co-authorship network is due to papers with three or more coauthors. Such papers introduce trivial triangles of collaborating authors, thereby increasing the clustering coefficient. 
This problem, that was raised by Newman et al. \cite{newman2}, was circumvented by studying directly the bipartite network, in order to infer the authors community structure. Newman et al. showed on some examples that these high order interactions may account for  one half of the clustering coefficient.
One should note, however, that if this approach offers a well-defined theoretical framework for bipartite networks, it suffers a lack of transparency as compared to the original projection method, i.e. it does not allow a clear visualisation of the unipartite structure.

In this article, we propose an alternative approach that is based on a more refine unipartite projection, and follows Statistical Mechanics usual expansion methods. To do so, we focus on a small dataset, retrieved from the arXiv database and composed of articles dedicated to complex network theory. This choice is motivated by their relatively few co-authors per article, a property typical to theoretical physics papers \cite{grossman}. 
Our method consists in discriminating the different kinds of scientists collaborations, based upon the number of co-authors per article. This discrimination leads  to a diagram representation \cite{feynman, mayer} of co-authorship (see also \cite{berg} for the applicability of Feynman diagrams in complex networks). The resulting N-body projection  reconciles the visual features of the usual projection,  and the exact description of Newman's theoretical approach. Empirical results confirm the importance of high order collaborations in the network structure. Therefore, we introduce in the last section a simple  network model, that is based on random triangular connections between the nodes. We study numerically percolation for the model.

\section{N-body projection method}
\begin{figure}
\includegraphics[angle=-90,width=3.50in]{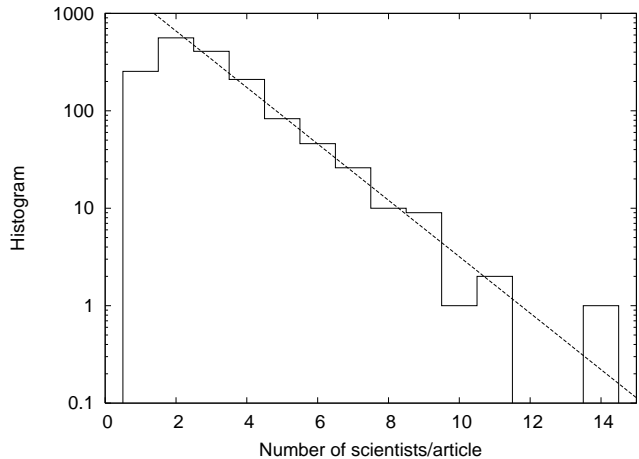}% Here is how to import EPS art
\caption{\label{dist} Histogram of the number of scientists/articles, $n$. The dashed line corresponds to the fit $e^{-\frac{n}{1.5}}$.}
\end{figure}

The data set contains all articles from arXiv in the time interval $[1995:2005]$, that contain the word {\em "network"}  in their abstract and are classified as {\em "cond-mat"}. In order to discriminate the authors and avoid spurious data, we checkeed the names and the first names of the authors. Moreover, in order to avoid multiple ways for an author to cosign a paper, we also took into account the initial notation of the prenames. For instance, {\em Marcel Ausloos} and {\em M. Ausloos} are the same person, while {\em Marcel Ausloos} and {\em Mike Ausloos} are considered to be different. Let us stress that this method may lead to ambiguities if an initial refers to two different first names, e.g. M. Ausloos might be Marcel or Mike Ausloos. Nonetheless, we have verified that this case occurs only once in the data set (Hawoong, Hyeong-Chai and H. Jeong), so that its effects are negligible. In that sole case, we attributed the papers of H. Jeong to the most prolific author (Hawoong Jeong in the dataset). Given this identification method, we find $n_P=2533$ persons and $n_A=1611$ articles.  The distribution of the number of co-authors per article  (Fig.\ref{dist}) shows clearly a rapid exponential decrease, associated to a clear predominance of small collaborations, as expected. 

Formally, the bipartite structure authors-papers may be mapped exactly on the vector of matrices $\mathcal{M} $ defined by:
\begin{equation}
\label{one}
\mathcal{M} = [{\bf M}^{(1)} , {\bf M}^{(2)}, ... ,  {\bf M}^{(j)} ,...., {\bf M}^{(n_P)}]
\end{equation}
where  ${\bf M}^{(j)}$ is a square $n_P^j$ matrix that accounts for all articles co-authored by $j$ scientists.  By definition, the element $M^{(j)}_{a_1 ... a_j}$ are equal to the number of  collaborations between the $j$ authors $a_1... a_j$.
In the following, we assume that co-authorship is not a directed relation, thereby neglecting the position of the authors in the collaboration, e.g. whether or not the author is the first author.
This implies that the matrices are symmetric under permutations of indices. Moreover, as people can not collaborate with themselves, the diagonal elements $M^{(j)}_{a a ... a}$ vanish by construction. For example, $M^{(1)}_{a_1}$ and $M^{(2)}_{a_1 a_2}$ represent respectively the total number of papers written by $a_1$ alone, and the total number of papers written by the pair ($a_1$, $a_2$). 

\begin{figure}
\hspace{0.4cm}
\includegraphics[width=3.30in]{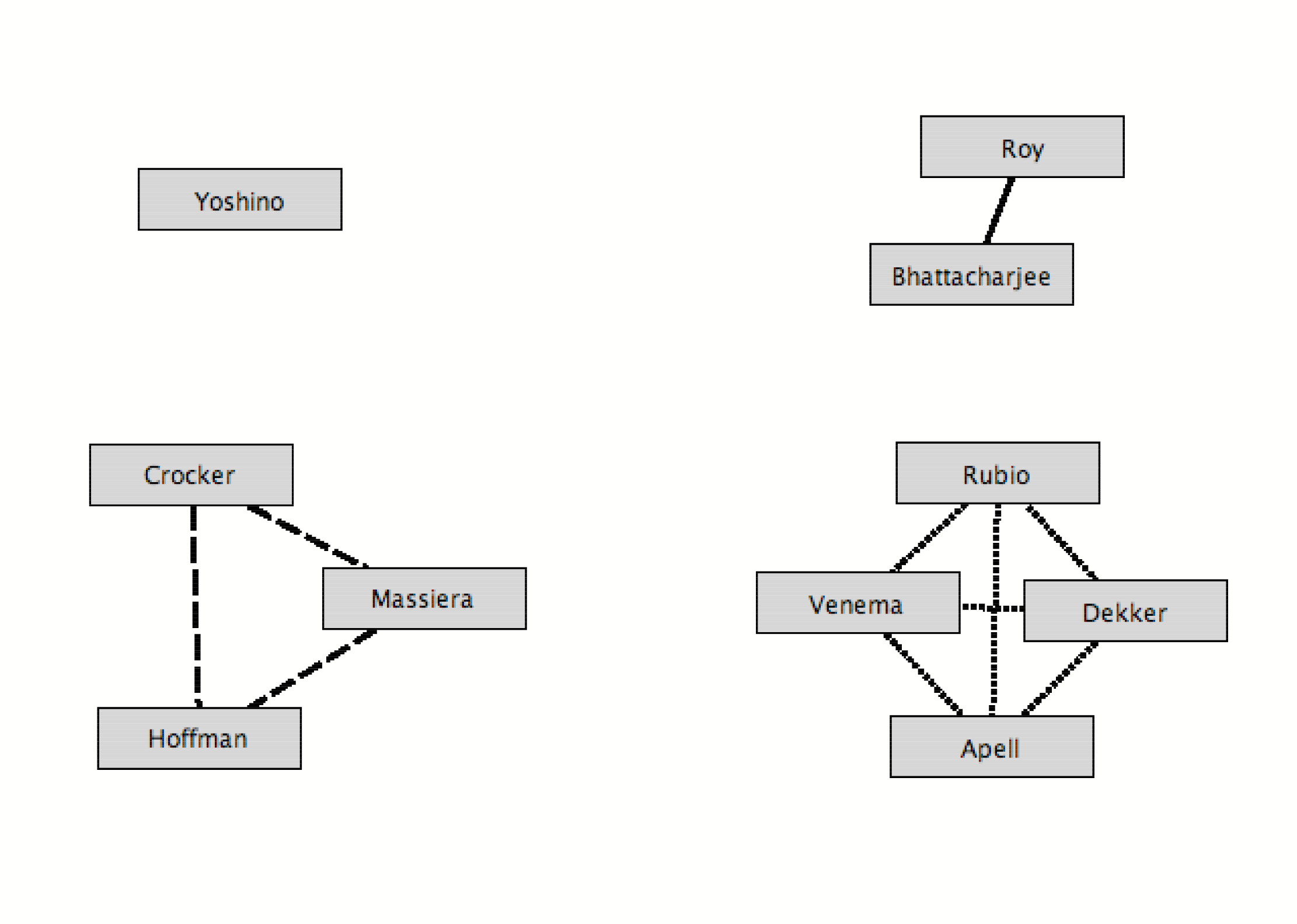}% Here is how to import EPS art
\caption{\label{fff} Graphical representation of the 4 most basic authors interactions, namely 1, 2, 3, 4 co-authorships. }
\end{figure}

\begin{figure}
\hspace{0.9cm}
\includegraphics[width=3.50in]{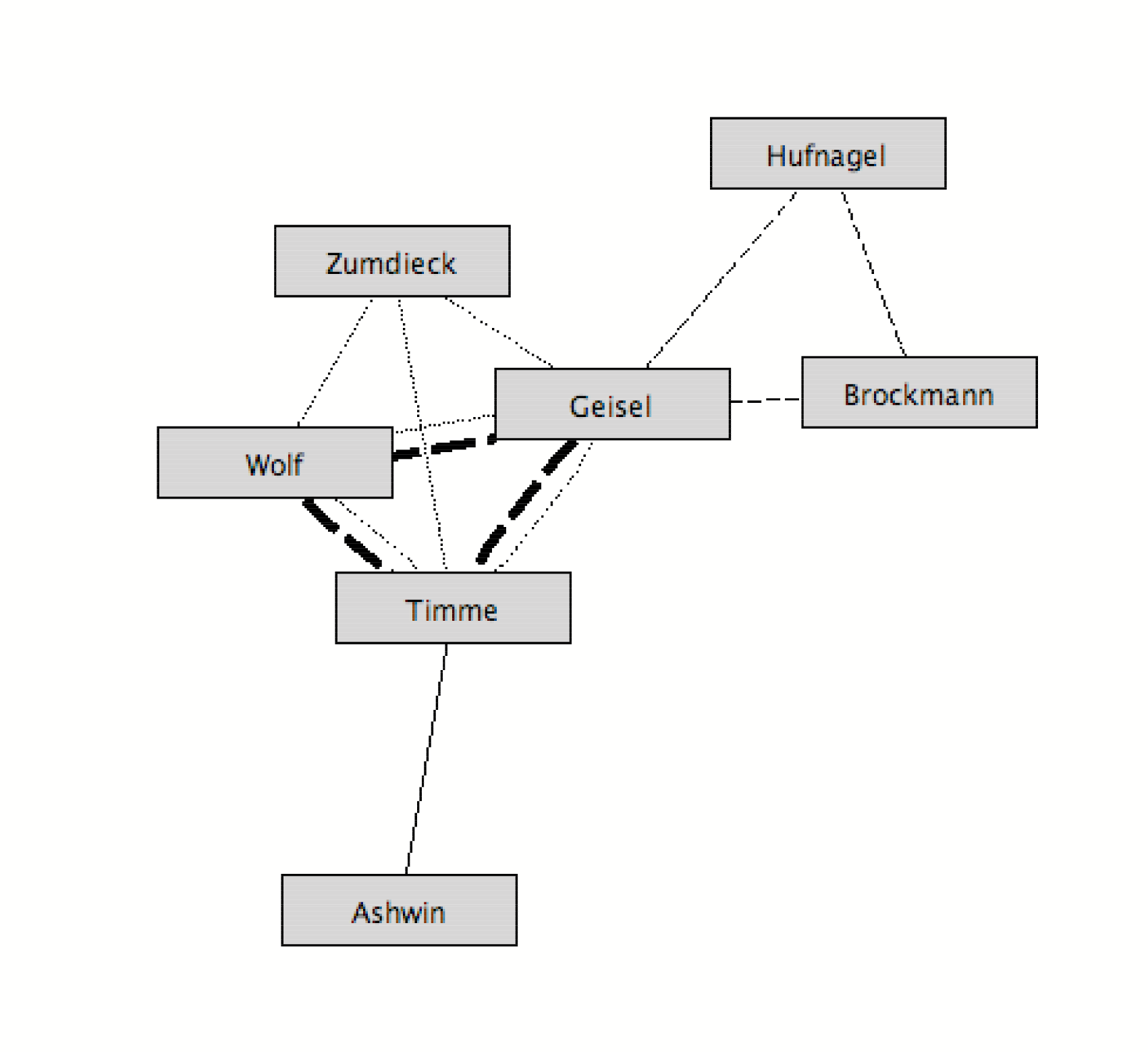}% Here is how to import EPS art
\caption{\label{example} Graphical representation of the co-authorship network. This small sub-network accounts for  1 two-authors collaboration,  {\em (Timme, Ashwin)}; 4 three-authors collaborations, 3 times {\em (Timme, Wolf, Geisel)} and  {\em (Geisel, Hufnagel, Brockmann))}; 1 four-authors collaboration {\em (Timme, Wolf, Geisel, Zumdieck)}. Because the triplet {\em (Timme, Wolf, Geisel)} collaborates three times, its links are three times larger than the other links.}
\end{figure}

A way to visualize $\mathcal{M}$ consists in a network whose nodes are the scientists, and whose links are discriminated by their shape. The intrinsic co-authorship interactions form loops (order 1), lines (order 2), triangles (order 3) (see Fig.\ref{fff})...
To represent the intensity of the multiplet interaction, the width of the lines is taken to be proportional to the number of collaborations of this multiplet.
Altogether, these rules lead to a graphical representation of $\mathcal{M}$, that is much more refine than the usual projection method  (Fig.\ref{example}).  

\begin{figure}
\hspace{2cm}
\includegraphics[width=3.50in]{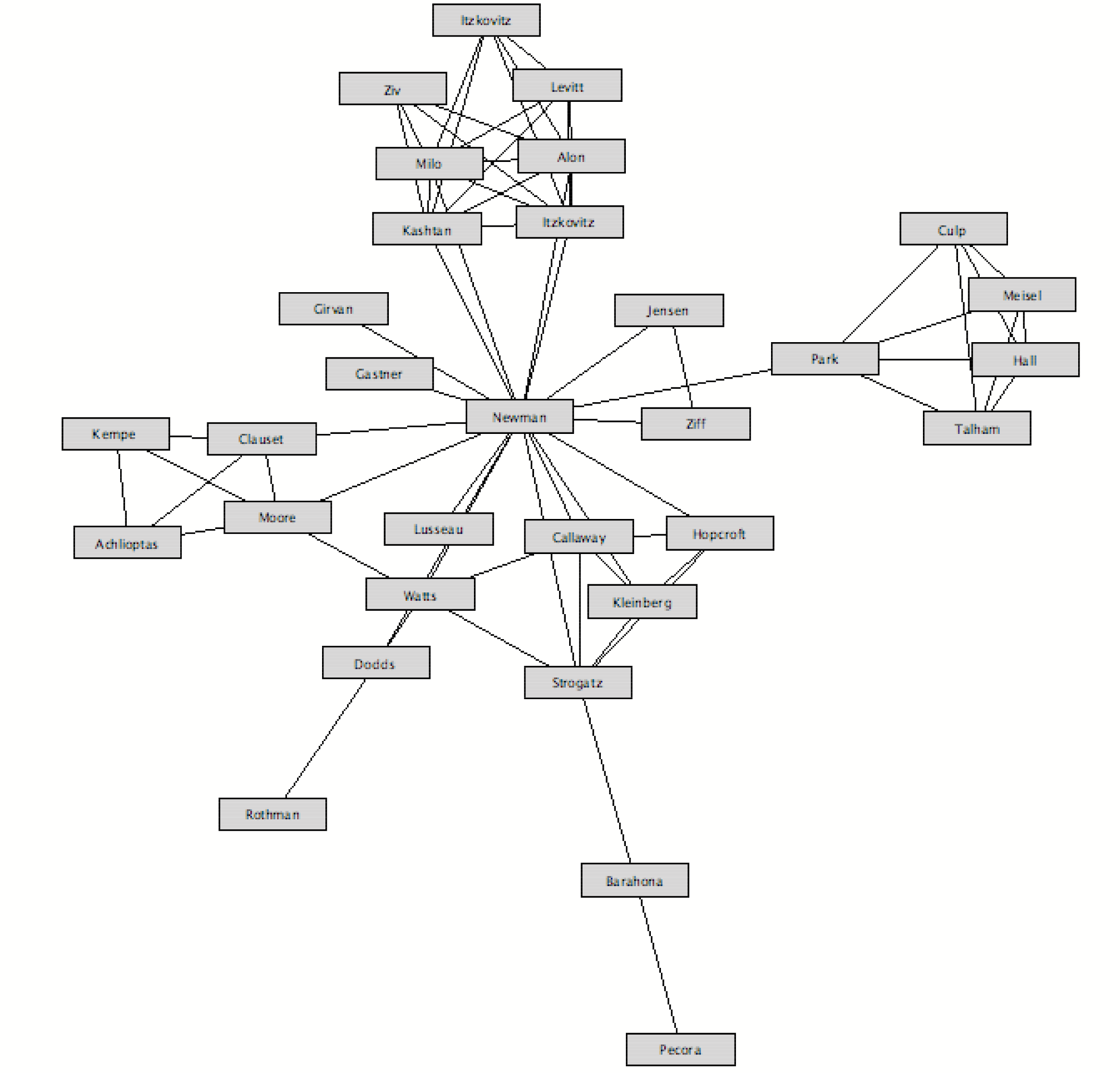}

\hspace{1cm}
\includegraphics[width=3.50in]{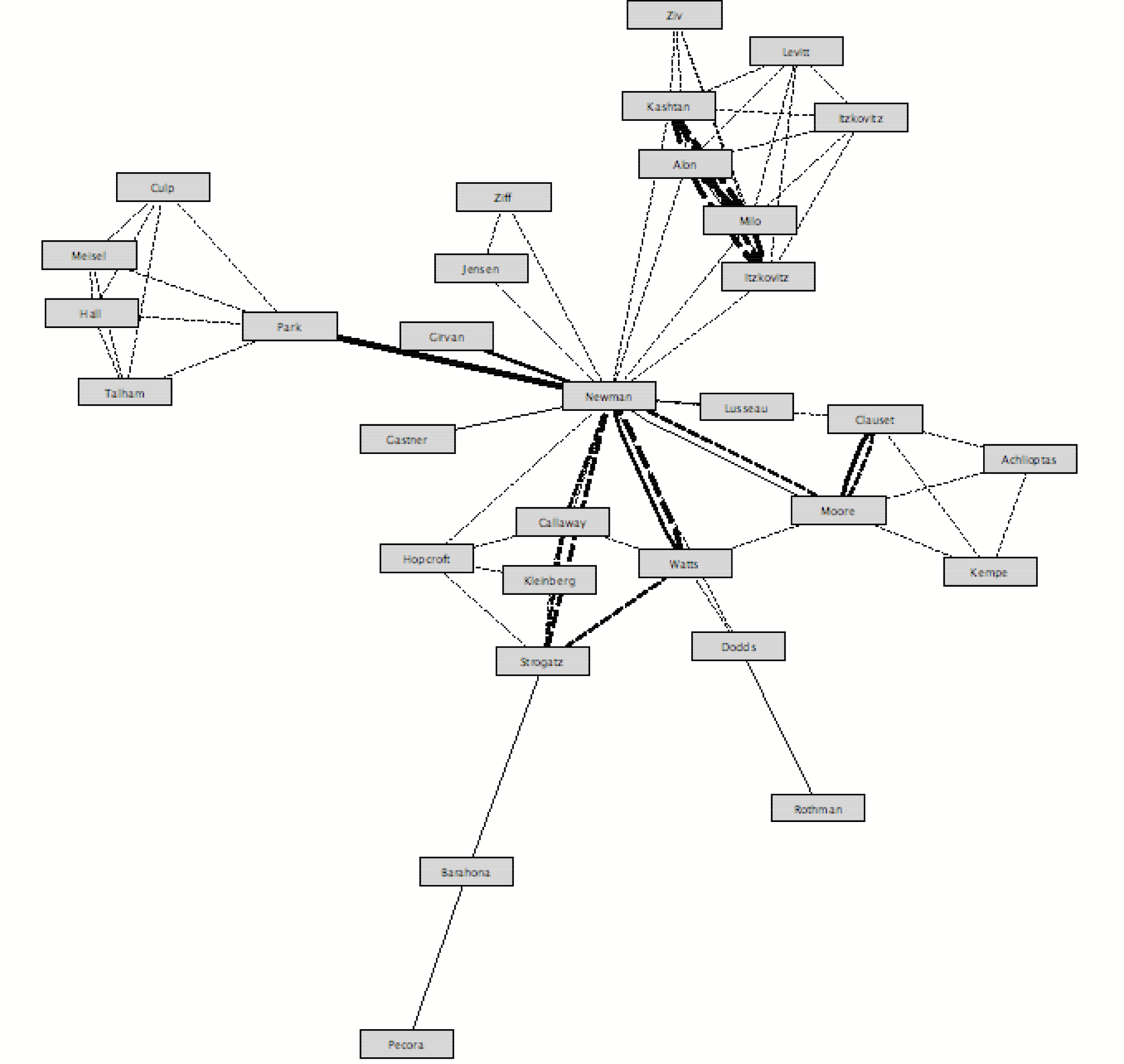}% Here is how to import EPS art
\caption{\label{basic} 3-body projection of the bipartite network. For the sake of clarity, we focus on a small sub-cluster, centered around the collaborations of M. Newman. The upper figure is the usual projection method \cite{newman2} . The lower figure is the triangular projection (\ref{three}) of the same bipartite network.}
\end{figure}

It is important to point out that the vector of matrices $\mathcal{M}$ describes without approximation the bipartite network, and that it reminds the Liouville distribution in phase space of a Hamiltonian system. Accordingly, a relevant macroscopic description of the system relies on a coarse-grained reduction of its internal variables.
The simplest reduced matrix is the one-scientist matrix ${\bf R}^{(1)}$ that is obtained by summing over the N-body connections, $N\geq 2$:
\begin{eqnarray}
R^{(1)}_{a_1} = M^{(1)}_{a1}  +  \sum_{a_2} M^{(2)}_{a_1 a_2} +  \sum_{a_2}\sum_{a_3<a_2} M^{(3)}_{a_1 a_2 a_3}+ .... \cr +  \sum_{a_2}.... \sum_{a_j<a_{j-1}} M^{(j)}_{a_1 ... a_j}+...
\end{eqnarray}
It is straightforward to show that the elements $R^{(1)}_{a_j}$ denote the total number of articles written by the scientist $a_j$.
The second order matrix:
\begin{eqnarray}
R^{(2)}_{a_1 a_2} = M^{(2)}_{a_1 a_2} +  \sum_{a_3} M^{(3)}_{a_1 ... a_3}+.... \cr +  \sum_{a_3}.... \sum_{a_j<a_{j-1}} M^{(j)}_{a_1 ... a_j}+...
\end{eqnarray}
Its elements represent the total number of articles written by the pair of scientists ($a_1$, $a_2$). 
Remarkably,  this matrix reproduces the usual projection method (Fig. \ref{projection}), and  obviously simplifies  the structure of the bipartite structure by hiding the effect of high order connections.
The three-scientist matrix read similarly:
\begin{eqnarray}
\label{three}
R^{(3)}_{a_1 a_2 a_3} = M^{(3)}_{a_1 a_2 a_3} +  \sum_{a_4} M^{(4)}_{a_1 ... a_4}+.... \cr +  \sum_{a_4}.... \sum_{a_j<a_{j-1}} M^{(j)}_{a_1 ... a_j}+...
\end{eqnarray}
This new matrix counts the number of papers co-written by the triplet ($a_1$, $a_2$, $ a_3$), and may be represented by a network whose links are triangles relating three authors. The generalization to higher order matrices ${\bf R}^{(j)}$ is straightforward, but, as in the case of the BBGKY hierarchy, a truncature of the vector $\mathcal{M}$ must be fixed at some level in order to describe usefully and compactly the system.
It is therefore important to point that the knowledge of ${\bf M}^{(2)}$ together with ${\bf R}^{(3)}$ is completely sufficient in order to characterize the triangular structure of $\mathcal{M}$. Consequently, in this paper, we stop the reduction procedure at the 3-body level, and define the triangular projection of $\mathcal{M}$ by the application:
\begin{eqnarray}
[M^{(1)}_{a1} , M^{(2)}_{a_1 a_2} ,  M^{(3)}_{a_1 a_2 a_3} ,...., M^{(n_P)}_{a_1... a_{n_P}}] \cr \rightarrow [M^{(1)}_{a1} , M^{(2)}_{a_1 a_2} ,  R^{(3)}_{a_1 a_2 a_3} ] 
\end{eqnarray}
The triangular projection is depicted in Fig. \ref{basic}, and compared to the usual projection method.
In order to test the relevance of this description, we have measured in the data set the total number of triangles generated by edges. We discriminate two kinds of triangles: those which arise from {\bf one} 3-body interaction of ${\bf R}^{(3)}$, and those which arise {\bf only} from an interplay of different interactions. There are respectively 5550 and 30 such triangles, namely $99.5 \%$ of triangles are of the first kind. This observation by itself therefore justifies the detailed projection method introduced in this section, and shows the importance  of co-authorship  links geometry in the characterization of network structures, precisely the clustering coefficient in the present case.

\section{Triangular Erd\"os-Renyi networks}

\begin{figure}
\includegraphics[width=2.50in]{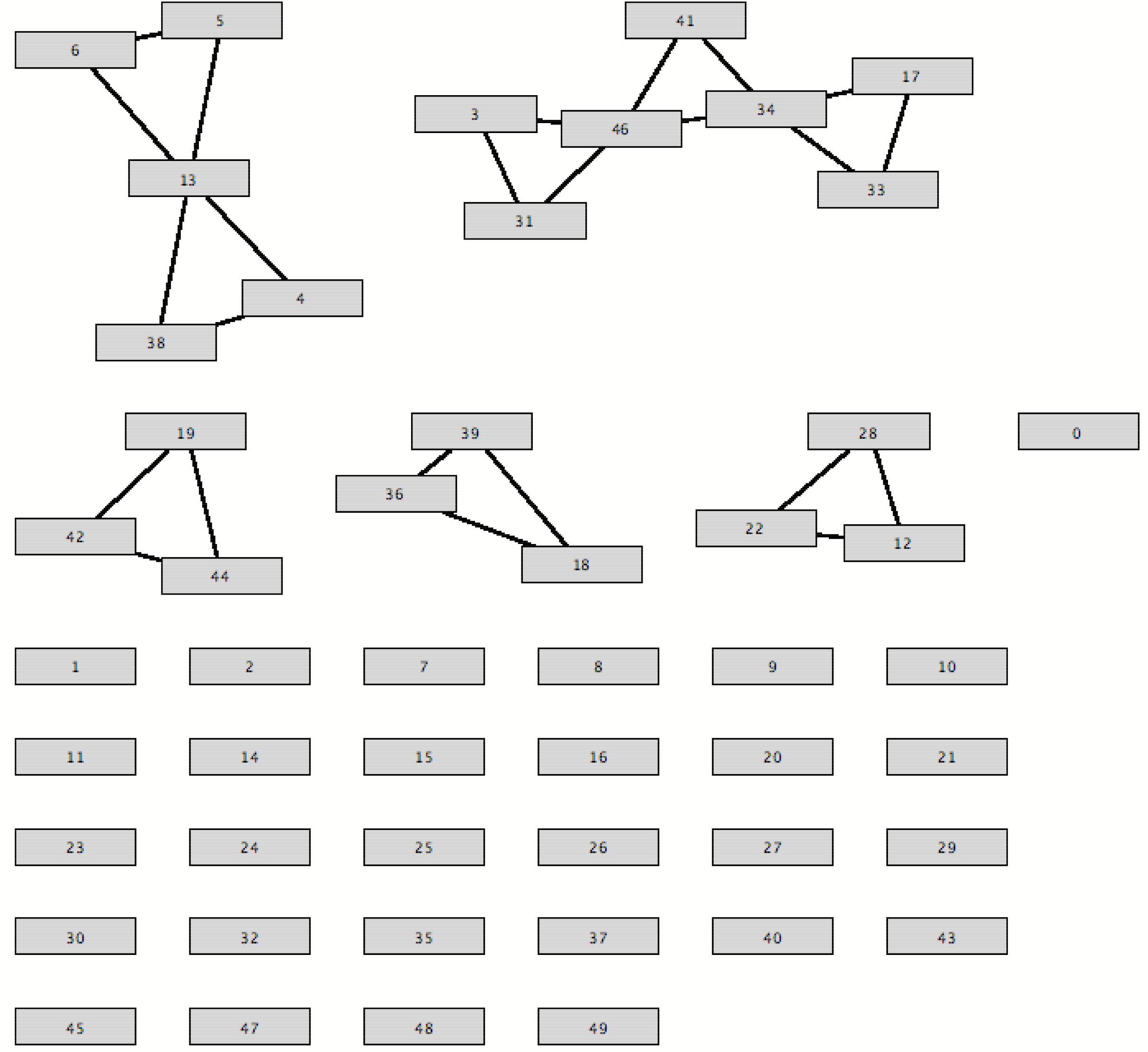}
\includegraphics[width=2.50in]{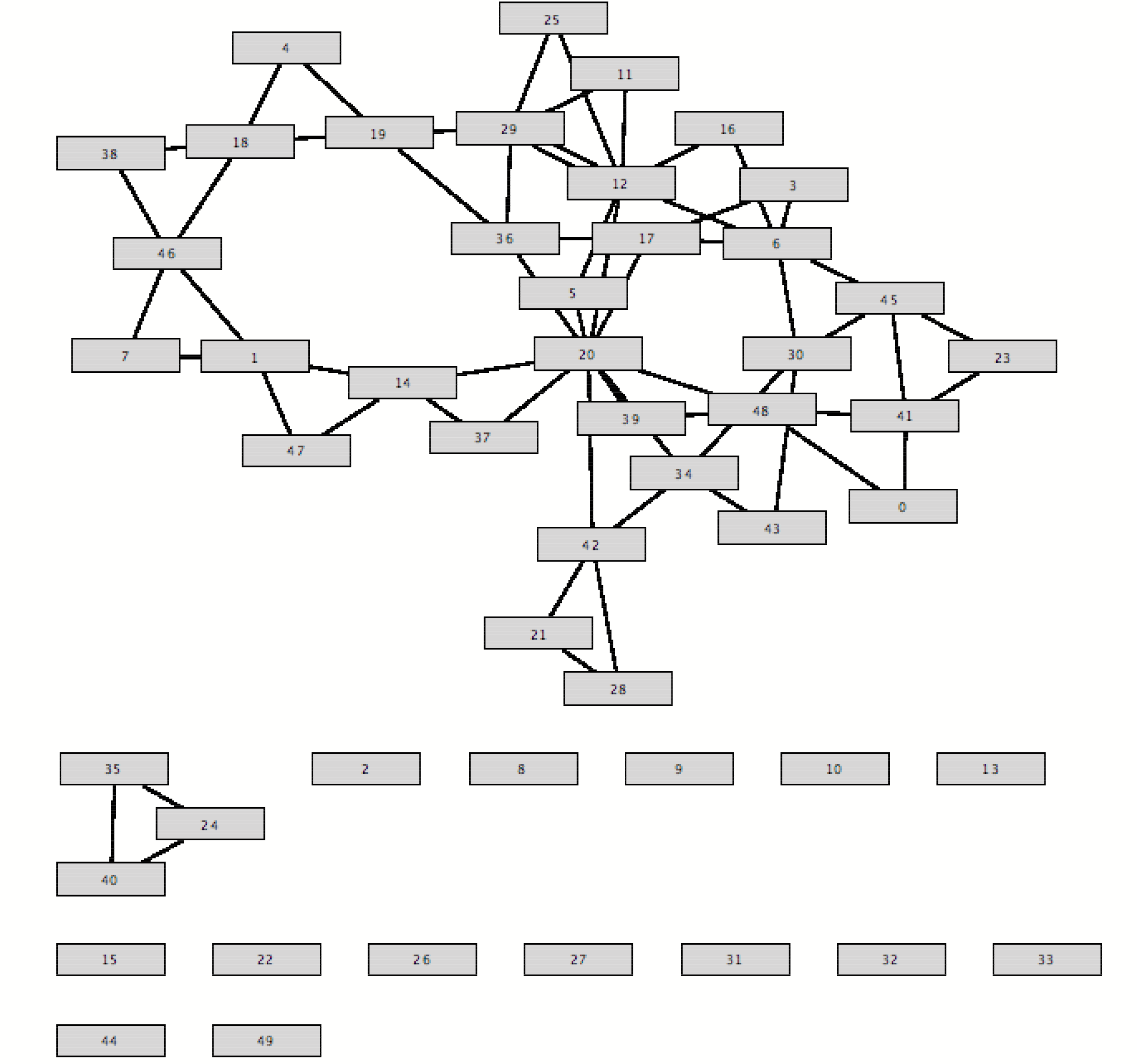}% Here is how to import EPS art
\caption{\label{perco} Percolation transition in the $\mbox{ERN}^3$ model with 50 nodes, from a dilute phase with small disconnected islands (8 triangles) to a percolated phase with one giant cluster (20 triangles).}
\end{figure}

The empirical results of the previous section have shown the significance of N-body connections in social networks. A more complete framework for networks is therefore required in order to describe correctly the system complexity.
In this article, we focus on the most simple generalization, namely a network whose links relate triplets of nodes.  To so, we base our modeling on the Erd\"os-Renyi  uncorrelated random graph \cite{renyi}, i.e. the usual prototype to be compared with more complex 
random graphs.
The usual Erd\"os-Renyi network (ERN) is composed by $N_n$ labeled nodes 
connected by $N_e^{(2)}$ edges, which are chosen randomly from 
the $N_n (N_n-1) /2$ possible edges. In this paper, we define the triangular ER network ($\mbox{ERN}^3$) to be
composed by $N_n$ labeled nodes, 
connected by $N_e^{(3)}$ triangles, which are chosen randomly from 
the $N_n (N_n-1) (N_n-2)/6$ possible triangles. As a result,  connections in the system relate triplets of nodes $(a_1, a_2, a_3)$, and the matrix vector $\mathcal{M}$ reduces to the matrix ${\bf M}^{(3)}$.
Before going further, let us point that the clustering coefficient of  triangular ER networks is very high by construction, but, contrary to intuition, it is different from 1 in general. For instance, for the two triplets $(a_1, a_2, a_3)$ and $(a_1, a_4, a_5)$, the local clustering coefficient of $a_1$ is equal to $\frac{1}{3}$.

\begin{figure}
\includegraphics[angle=-90,width=3.50in]{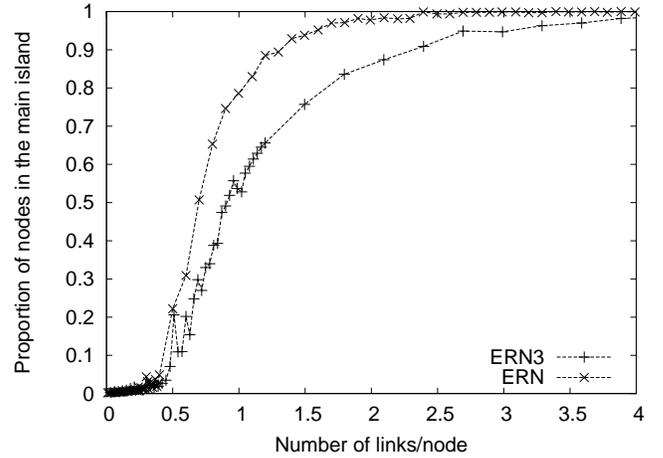}
\caption{\label{transition} Proportion of nodes in the main island, as a function of the number of links/node, in the ERN and the $\mbox{ERN}^3$ model.}
\end{figure}

In this paper, we focus numerically on the percolation transition \cite{vicsek} in $\mbox{ERN}^3$, i.e. on the appearance of a giant component by increasing the number of nodes in the system (Fig.\ref{perco}). This transition is usually associated to dramatic changes in the  topological structure, that are crucial to ensure communicability between network nodes, e.g. the spreading of scientific knowledge in the case under study. In the following, we work at fixed number of nodes, and focus on the proportion of nodes in the main cluster as a function of the number of binary links in the system.
Moreover, in order to compare results with the usual ERN, we do not count twice redundant links, i.e. couples of authors who interact in different triplets. For instance, the triplet $(a_1, a_2, a_3)$ accounts for 3 binary links, but $(a_1, a_2, a_3)$ and $(a_1, a_2, a_4)$ account together for 5 links, so that $N_e^{(3)} \neq 3 N_e^{(2)}$ in general. Whatever, this detailed counting has small effects on the location of the percolation transition. Numerical results are depicted in figure \ref{transition}, where we consider networks with $N_n=1000$. Obviously, the triangular structure of interactions displaces the bifurcation point, by requiring more links in order to observe the percolation transition. This feature comes from the triangular structure of connections that restrains the network exploration as compared to random structures. Indeed,  3 links relate only 3 nodes in $\mbox{ERN}^3$, while 3 links typically relate 4 nodes in ERN. Finally, let us stress that the same mechanism takes place in systems with high clustering coefficients \cite{clustering, preparation}.

 \section{Conclusion}
In this paper, we show the importance of N-body interactions in co-authorships networks. By focusing on data sets extracted from the arXiv database, we introduce a  way to project bipartite networks onto unipartite networks. This approach generalizes usual projection methods by accounting for the complex geometrical figures connecting authors. To do so, we present a simple theoretical framework, and define N-body reduced and projected networks. The graphical representation of these simplified networks rests on a "shape-based" discrimination of the different co-authorship interactions (for a "color-based" version, see the author's website \cite{website}), and allows a clear visualization of the different mechanisms occurring in the system. Finally, we apply the method to some arXiv data subset, thereby showing the importance of such "high order corrections" in order to characterize the community structure of scientists. 
The empirical results motivate therefore a better study of networks with complex weighted geometrical links. In the last section, we focus on the simplest case by introducing a triangular random model, $\mbox{ERN}^3$.
Moreover, we restrict the scope by analyzing  the effect of the 3-body connection on percolation. A complete study of the topological of $\mbox{ERN}^3$ as well as its generalization to higher order connections is let for a forthcoming work.  

%\end{center}

{\bf Acknowledgements}

Figures \ref{fff}, \ref{example}, \ref{basic} and \ref{perco} were plotted thanks to the {\em visone} graphical tools.
This work 
has been supported by European Commission Project 
CREEN FP6-2003-NEST-Path-012864.

\end{document}